\documentclass[12pt]{article}
\usepackage[pctex32]{graphics}
\begin{document}
\vspace{2cm}
\begin{center}
~\\
~\\
{\bf  \Large Electric/Magnetic Field Deformed Giant Gravitons in Melvin Geometry}
\vspace{1cm}

                      Wung-Hong Huang\\
                       Department of Physics\\
                       National Cheng Kung University\\
                       Tainan,Taiwan\\

\end{center}
\vspace{1cm}
\begin{center}{\bf  \Large ABSTRACT } \end{center}
The rotating D3-brane in the $AdS_5 \times S^5$ spacetime could be blowed up to the spherical BPS configuration which has the same energy and quantum number of the point-like graviton and is called as a giant graviton.  The configuration is stable only if its angular momentum was less than a critical value of $P_c$. In this paper we investigate the properties of the giant graviton in the electric/magnetic Melvin geometries of deformed  $AdS_5 \times S^5$ spacetime which was obtained in our previous paper (hep-th/0512117, Phys. Rev. D73 (2006) 026007).  We find that in the magnetic Melvin spacetime the giant graviton has lower energy than the point-like graviton.  Also, the critical value of the angular momentum is an increasing function of the magnetic field flux $B$. In particular, it is seen that while increasing the angular momentum the radius of giant graviton  is initially an increasing function, then, after it reach its maximum value it becomes a decreasing function of the angular momentum.  During these regions the giant graviton is still a stable configuration, contrast to that in the undeformed theory.   Finally, beyond the critical value of angular momentum the giant graviton has higher energy than the point-like graviton and it eventually becomes unstable.   Our analyses show that the electric Melvin field will always render the giant graviton unstable.
\vspace{2cm}
\begin{flushleft}
E-mail:  whhwung@mail.ncku.edu.tw\\
\end{flushleft}


\newpage
\section{Introduction}
McGreevy, Susskind and Toumbas [1] found that a rotating D3-brane in the $AdS_5 \times S^5$ spacetime could expand into a stable configuration which has exactly the same quantum numbers as the point-like graviton and is thus called as a giant graviton.   This expanded brane wraps the spherical part of the $S^5$ spacetime and is stabilized against shrinking by the Ramond-Ramond (RR) gauge field.  The size of the giant graviton increase with its angular momentum and, as the radius of the brane cannot be greater than the radius of the spacetime, there exists an upper bound for the angular momentum of the brane. This fact realizes the stringy exclusion principle [2].   The authors in [3] had proved that the giant gravitons are BPS configurations which preserve the same supersymmetry as the point-like graviton.   As the giant graviton has exactly the same quantum numbers as the point-like graviton they can tunnel into each others.

  It was also shown in [3] that there exist ``dual '' giant graviton consisting of spherical brane expanding into the AdS part of the spacetime, which, however, do not have an upper bound on their angular momentum due to the non-compact nature of the AdS spacetime.   While the giant graviton could tunnel into the trivial point-like graviton the investigations had shown that there is no direct tunneling between the giant graviton and its dual counterpart in AdS [3-6]. The microscopical description of giant gravitons had also been investigated in [7,8].  The blowing up of gravitons into branes can take place in backgrounds different from AdS $\times$ S.  In [9] it was found that there are giant graviton configurations of D(6-p)-branes moving in the near-horizon geometry of a dilatonic background created by a stack of Dp-branes. The giant graviton solutions for probes which move in the geometry created by a stack of non-threshold bound states of the type (D(p-2), Dp) had also been found in [10].

  The giant graviton in the background B-field had been studied in [11] and properties of the non-spherical giant graviton have been considered in [12].  The problems of giant graviton solutions in Frolov's three parameter generalization of the Lunin-Maldacena background [13] were investigated in a recent paper [14].  These investigations have shown that the deformations have inclinations to make the giant graviton unstable.  

   In this paper we will investigate the effects of the external electric/magnetic fields on the giant gravitons.   Using the electric/magnetic Melvin spacetimes which was found in our previous paper [15] we will first show that in the magnetic Melvin spacetime the magnetic field deformed giant graviton has lower energy than the point-like graviton.   Thus the magnetic Melvin field has effect to stabilize the giant graviton and to suppress it from tunneling into the point-like graviton.  In particular, we have seen that while increasing the angular momentum of giant graviton the radius of the giant graviton is initially an increasing function, the property is the same as that in the undeformed giant graviton.   However, after it reach its maximum value the radius of the giant graviton becomes a decreasing function of the angular momentum.  During these regions the giant graviton is still a stable configuration, contrast to that in the undeformed theory.   Finally, beyond a critical value of angular momentum $P_c$ the configuration has higher energy than the point-like graviton and giant graviton  eventually becomes unstable.   The critical value of the angular momentum $P_c$ therein is an increasing function of the magnetic field flux $B$.   We have also studied the giant graviton in the electric Melvin spacetime.  The results show that the electric Melvin field always render the giant graviton unstable.

  Note that our previous investigation [16] showed that beyond a sufficient high temperature the giant graviton is more stable than a trivial configuration.  This paper also shows that the magnetic Melvin field has effect to stabilize the giant graviton and to suppress it from tunneling into the trivial point-like graviton.  

 Note also that Giant graviton, being a BPS object, provides a natural framework for the study of the gauge /gravity correspondence in supersymmetric examples [17].   After studying the zero coupling limit of N = 4 super Yang-Mills theory the candidate operators dual to giant gravitons had been proposed in [18,19]. Recently there was a progress in constructing non-supersymmetric examples of the gauge/gravity correspondence, as Lunin and Maldacena [13] had demonstrated that certain deformation of the $AdS_{5}\times S^{5}$ background corresponds to a $\beta$-deformation of $N=4$ SYM gauge theory in which the supersymmetry being broken \footnote{More references concerning this issues could be found in [15].}.   As  the supersymmetry is broken by the magnetic or electric field, the deformed giant graviton in the electric/magnetic Melvin geometries could provide us the systems to investigate the non-supersymmetric examples of the gauge/gravity correspondence. Therefore  it is important to find the Yang-Mills operators corresponding to the  electric/magnetic field deformed giant graviton, as those analyzed in [14] for the giant graviton in the Frolov's three parameter generalization of the Lunin-Maldacena background [14].  Also, the correspondence between the deformed giant graviton on the electric/magnetic field Melvin geometries and the spin chain in the deformed geometries [20], which can improve our understanding of the AdS/CFT correspondence, are deserved to be investigated.  These important problems are left to further research.  

   In section 2 we investigate the giant graviton in magnetic Melvin spacetime and in section 3 we investigate the giant graviton in  electric Melvin spacetime.  In section 4 we discuss our results.

  For convenience, let us briefly describe the method to obtain the deformed $AdS_5 \times S^5$ background [15].   The full $N$ M2-branes solution is given by [21]
$$ds^2_{11}=H^{-2\over3}\left(-dt^2+dx_1^2+dx_2^2\right)+H^{1\over3}\left
(dx_3^2 + d\rho^2+\rho^2 \Omega_5^2+dx_{11}^2\right),$$
$$ ~~~~~~~~~~d\Omega_5^2 \equiv d\theta^2 + \cos^2\theta d\phi^2 + \sin^2\theta\left(d\chi_1^2+\sin^2\chi_1 \left(d\chi_2^2+\sin^2\chi_2d\chi^2_3\right)\right), \eqno{(1.1)}$$
$$A^{(3)} = H^{-1} dt\wedge dx_1\wedge dx_2.\eqno{(1.2)}$$
$H$ is the harmonic function defined by 
$$ H = 1+ {L\over r^{D-p-3}}, ~~~~~~~r^2\equiv x_3^2+ \rho^2+x_{11}^2 , 
~~~~L \equiv {16\pi G_D\,T_p \,N\over D-p-3 },\eqno{(1.3)}$$
in which  $G_D$ is the D-dimensional Newton's constant and $T_p$ the p-brane tension. In the case of (1.1), $D=11$ and $p=2$.   We can transform the angle $\phi$ by mixing it with the compactified  coordinate $x_{11}$ in the following substituting 
$$\phi \rightarrow \phi + B x_{11},\eqno{(1.4)}$$ 
to obtain a magnetic Melvin flux  [22] or transform the time $t$ by mixing it with the compactified coordinate $x_{11}$ in the following substituting
$$ t  \rightarrow t - E x_{11},\eqno{(1.5)}$$ 
to obtain an electric Melvin flux [23].  Using the above substitutions the line element can be expressed as 
$$ds_{11}^2= e^{-2\Phi/3}ds_{10}^2+  e^{4\Phi/3} (dx_{11}+2 A_\mu dx^\mu )^2 , \eqno{(1.6)} $$
in which $\Phi$ is the dilaton field, $A_\mu $ the 1-form potential and $ds_{10}^2$ is the line element of 10D IIA background.  In this decomposition into ten-dimensional fields which do not depend on the $x_{11}$ the ten-dimensional Lagrangian density becomes 
$$ {\cal L}_{10D} = {\cal R}- 2 (\nabla \Phi)^2 - e^{2\sqrt 3 \Phi}~ F_{\mu\nu}F^{\mu\nu},\eqno{(1.7)}$$
in which $F_{\mu\nu}$ is the EM field strength.  To find the spacetime of a stack of N D3-branes we now perform the T-duality transformation [24] on the coordinate $x_{3}$.  After taking the near-horizon limit we then obtain the magnetic field deformed $AdS_5 \times S^5$ background in (2.1) and electric field deformed $AdS_5 \times S^5$ background in (3.1), as detailed in [15]. 

\section{Giant Graviton in Magnetic Melvin Spacetime}
The background of  magnetic Melvin spacetimes found in [15] is described by 
$$ds_{10}^2 = L \sqrt{1+B^2L^2Z^{-2}\cos^2\theta}\left[{1\over Z^2}(- dt^2+ dx_1^2+dx_2^2 +{1\over \sqrt{1+B^2L^2Z^{-2}}}dx_3^2+dZ^2) +\right.\hspace{2cm} $$
$$\left.\hspace{1cm} \left(d\theta^2 + {\cos^2\theta d\phi^2\over 1+ B^2L^2Z^{-2}\cos^2\theta} + \sin^2\theta\left(d\chi_1^2+\sin^2\chi_1 \left(d\chi_2^2+\sin^2\chi_2d\chi^2_3\right)\right)\right)\right],\eqno{(2.1)}$$ 
$$e^{\Phi} = \sqrt{1+B^2\,\cos^2\theta},\eqno{(2.2)}$$
$$A_\phi = {B\,\cos^2\theta\over 2 \left(1+B^2\,\cos^2\theta\right)}.\eqno{(2.3)}$$
in which we define $Z\equiv L^2/\rho$.  $\Phi$ is the corresponding dilaton field and $A_\phi$ is called as a magnetic Melvin field [22].    In the case of  $B=0$  the above spacetime becomes the well-known geometry of $AdS_5\times S^5$. Thus, the background describes the magnetic-field deformed $AdS_5\times S^5$.  

   The rotating giant graviton we will search is the D3-brane wrapping the spherical $\chi_i$ spacetime.   The world-volume coordinate $\sigma_\mu$ are identified with the space-time coordinates by [1,3]
$$\sigma_0= t,~~\sigma_1 = \chi_1,~~\sigma_2 = \chi_1,~~\sigma_3 = \chi_1,\eqno{(2.4a)}$$
and 
$$\phi=\phi(t).\eqno{(2.4b)}$$
The giant graviton will be fixed on the spatial coordinates at  $x_1=x_2=x_3=0$ with a fixed value of $Z$.  As we merely want to see the effect of magnetic Melvin filed on the giant graviton, we will let $L=\rho=1$ for convenience.   This means that we have used the scale  $ {B^2 L^2 Z^{-2}} \rightarrow  B^2$.   

The 4-form RR potential on the deformed $S^5$ is 
$$A^{(4)}_{\phi\chi_1\chi_2\chi_3} = \sqrt{g_\chi}~A(\theta),\eqno{(2.5a)}$$
where  $\sqrt{g_\chi}= \sin^2\chi_1~\sin\chi_2$ and 
$$A(\theta) = {8\over 15B^4}\left(\left(1+B^2\right)^{5/2}-\left(1+B^2\,\cos^2\theta\right)^{5/2}\right)-{4\over 3B^2}\left(1+B^2\,\cos^2\theta\right)^{3/2}\sin^2\theta.\eqno{(2.5b)}$$
Above definition of 4-form RR potential has the property that 
$$F_{\theta\phi\chi_1\chi_2\chi_3} = \sqrt{1+B^2\cos^2\theta}~\cos\theta\,\sin^3\theta \sqrt{g_\chi},\eqno{(2.6)}$$
and $A^{(4)}\rightarrow \sin^4\theta$ as $B \rightarrow 0$, which is that used in the undeformed $S^5$ theory [1,3].  (Notice that the corresponding radius of $S^5$ in [1] is taken to be $L =1$ in our notation.)

To proceed, we know that the D3-brane action may be written as
$$S= -\int d^{4}\sigma\ \sqrt{-g} +  \int P[A^{(4)}],\eqno{(2.7)}$$
where $g_{ij}$ is the pull-back of the spacetime metric to the
world-volume, and $P[A^{(4)}]$ denotes the pull-back of the $4$-form potential.    Now we can use the classical rotating D3-brane solution ansatz  (2.4) to find the associated Lagrangian  
$${\cal L}= - \sqrt{1+B^2\,\cos^2\theta -\cos^2\theta~\dot\phi^2}~\sin^3\theta + A(\theta)~\dot\phi. \eqno{(2.8)}$$
After the calculations the momentum conjugates to $\phi$ becomes
$$P={\cos^2\theta \sin^3\theta~\dot\phi\over\sqrt{1+B^2\,\cos^2\theta -\cos^2\theta~\dot\phi^2}} + A(\theta),\eqno{(2.9)}$$
and associated energy of the deformed giant graviton is 
$$H = {\sqrt{1+B^2\,\cos^2\theta }\over \cos\theta}  \sqrt{\left(P -A(\theta)\right)^2+ \cos^2\theta~\sin^6\theta},\eqno{(2.10)}$$
which becomes that in the undeformed system in the limit of $B =0$.  Note that the radius of the giant graviton, $R$, is the value $\sin\theta$ in our notation.

  Before using (2.10) to analyze the properties of the deformed giant graviton it is useful to know that the radius of an undeformed giant graviton is equal to its angular momentum, i.e. $ R=P$ [1].  Therefore, increasing the angular momentum of the undeformed giant graviton will increasing its size. In this case the giant graviton has same energy as the point-like graviton.  However, once its angular momentum is larger than 1, i.e. $P >1$, the configuration will has higher energy than that of the point-like graviton and giant graviton becomes unstable.   With these properties in mind we now use the above formula to plot the energy of the deformed giant graviton as a function of its radius $R$ with various angular momentum $P$, under a fixed magnetic flux $B=4$.   The results are shown in figure 1.  The figure has properties which are different form those in the undeformed case:

   1. The giant graviton has lower energy than that of the point-like graviton.   This means that the magnetic Melvin field will stabilize the giant graviton and suppress it from tunneling into the point-like graviton.

   2. Form the figure 1 we see that the stable radius $R$ of the giant graviton with angular momentum $P$ are at the values of $(P,R) \approx (1,0.3), (2.5, 0.99), (6, 0.7)$.  The numerical results show that while increasing the angular momentum of giant graviton the radius of the giant graviton is initially an increasing function, the property is the same as that in the undeformed giant graviton.   However, after it reach its maximum value  $R=1$ as $P\approx 2.6$ the radius of the giant graviton becomes a decreasing function of the angular momentum.   In these cases the giant graviton is still a stable configuration. 
\\

\hfil\scalebox{1}{\includegraphics{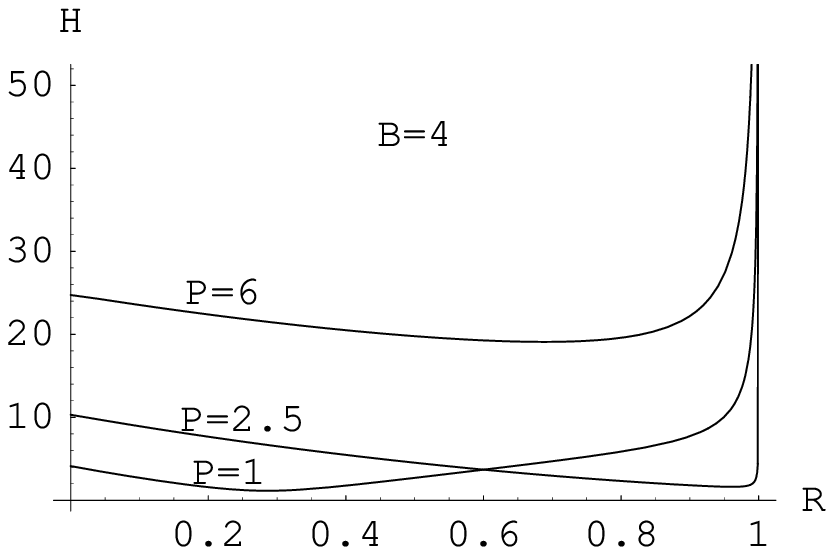}}\hfil\\
{\it Figure 1:  Energy of the giant graviton with various angular momentum P=1, 2.5, or 6 as a function of its radius $R$ under a magnetic flux $B=4$. Giant graviton has its maximum radius  $R=1$ as $P\approx 2.6$ and becomes unstable as its angular momentum $P > 30$.}
\\

  3. The numerical results show that in the case with a fixed magnetic flux $B=4$, the giant graviton with angular momentum $P> P_c \approx 30$  will have higher energy than the point-like graviton and it will become unstable.  The critical value of the angular momentum $P_c$ is found to be an increasing function of the magnetic field flux $B$.

 4. It is showd in [1,3] that the radius of the undeformed giant graviton with angular momentum $1< P < 1.125$  is a decreasing function of $P$ and giant graviton is unstable.  Our numerical results show that  the radius of the magnetic-field (B=4) deformed giant graviton with angular momentum $2.6< P < 30$  is a decreasing function of $P$ and, however, the deformed giant graviton is stable.   The magnetic field therefore could stabilize the deformed giant graviton in the region.

   It is interesting to see the properties in analyzing the case of small magnetic flux, $B \ll 1$.   In this limit the energy of the giant graviton (2.10) has a simple form
$$H\approx {\sqrt{P^2-2P\sin^4\theta+\sin^6\theta}\over \cos\theta} + 
{B^2\over \cos\theta}{3P^2 \cos^2\theta + 3 \sin^6\theta - 2\sin^{10}\theta- P (9-8\sin^2\theta)\sin^4\theta \over \sqrt{P^2-2P\sin^4\theta+\sin^6\theta}} .\eqno{(2.11)}$$
The stable radius calculated from the above equation is
$$R_{gg}^2 \equiv \sin^2\theta \approx P + {B^2\over 6}\left(9P -18P^2 + 8P^3\right).\eqno{(2.12)}$$
The energy of the above giant graviton $H(R_{gg})$ and that of the point giant graviton $H(0)$ calculated from (2.11) have the following relation
$$H(R_{gg}) -H(0) \approx {B^2\over 3} P^2(P-3),\eqno{(2.13)}$$
in which 
$$H(0) \approx P(1+ 3B^2),\hspace{1cm}H(R_{gg})\approx P(1+ 3B^2)+ {B^2\over 3} P^2(P-3).\eqno{(2.14)}$$
Using above two equations we can easily see the following properties. 

1.  Eq. (2.12) tells us that while increasing the angular momentum $P$ the radius of the giant graviton, $R_{gg}$, is initially an increasing function.  However, after it reach its maximum value, $R_{gg}=1$, the radius of the giant graviton becomes a decreasing function of the angular momentum.  Within this region, while the undeformed giant graviton is unstable as analyzed in [1,3],  the deformed giant graviton is still a stable configuration as analyzed in following. 

2.  As shown in [1] that the configurations with $1<P<P_M \equiv 1.125$ is the unstable giant graviton, therefore, to the first order of  the  magnetic flux $B^2$, we can substituting the value $P=P_M $ into eq. (2.13) and see that the deformed giant graviton has less energy than a point-like graviton.  Thus the magnetic Melvin field has effect to stabilize the unstable giant graviton and it will suppress the giant graviton from tunneling into the point-like graviton.

3. Eq.(2.13) tells us that giant graviton with angular momentum $P<3$ will have less energy than the point-like graviton, however, a prior assumption therein is the existence of the giant graviton.  In fact, numerical analyses have shown that beyond a critical value of angular momentum $P_c$ the giant graviton does not exist.

4.  In the undeformed case, the maximum radius is $R=1$ (Note that $R= \sin\theta$ in our notation.) which is found at a giant graviton with angular momentum $P_0=1$ [1].  Therefore, to the first order of $B^2$ field we can substitute $P=1$ into the correction in (2.12) to solve the equation $R=1 = P_m + {B^2\over 6}\left(9 -18 + 8\right)$ and find that  $P_m = 1+ {B^2\over 6}>1=P_0$.  Thus the angular momentum of a giant graviton with maximum radius in the magnetic Melvin background is larger than that in the undeformed system.

5. From (2.15) we see that the correction to the energy of magnetic-field deformed giant graviton is positive thus, from the AdS/CFT point of view [17-19], the correction of the anomalous dimensions of  operators in the dual SYM theory will be positive. 

\section{Giant Graviton in Electric Melvin Spacetime}
The background of  electric Melvin spacetimes found in [15] is described by 
$$ds_{10}^2=-{\rho^2/L\over\sqrt{1-{E^2\rho^4/ L}}}\left(dt^2 - dx_3^2\right) +\sqrt{1-{E^2\rho^4/ L}}\left[{\rho^2\over L} \left(dx_1^2+dx_2^2\right) + {L\over \rho^2} \left(d\rho^2+\rho^2 d\Omega_5^2\right)\right], \eqno{(3.1)}$$ 
$$e^{\Phi} =\sqrt{1-{E^2\rho^4/ L}}, \eqno{(3.2)}$$ 
$$A_t = {E^2{\rho^4/ L}\over 1-{E^2\rho^4/ L}}~~~~, \eqno{(3.3)}$$ 
in which
$$d\Omega_5^2 = d\theta^2 + \cos^2\theta d\phi^2 + \sin^2\theta\left(d\chi_1^2+\sin^2\chi_1 \left(d\chi_2^2+\sin^2\chi_2d\chi^2_3\right)\right),\eqno{(3.4)}$$
$\Phi$ is the corresponding dilaton field and $A_t$ is called as an electric Melvin field [23].   In the case of  $E=0$  the above spacetime becomes the well-known geometry of $AdS_5\times S^5$. Thus, the background describes the electric-field deformed $AdS_5\times S^5$.  

   We will consider a giant graviton configuration as that described in equation (2.4)  which is fixed on the spatial coordinates at  $x_1=x_2=x_3=0$.   As we merely want to see the effect of electric Melvin filed on the giant graviton, we will let $L=\rho=1$ for convenience.   This means that we have used the scale  $ {E^2 \rho^4\over L} \rightarrow  E^2$.  The 4-form RR potential on the deformed $S^5$ is 
$$A^{(4)}_{\phi\chi_1\chi_2\chi_3} = (1-E^2) \sin^4\theta \sqrt{g_\chi},\eqno{(3.5)}$$

    Now, through the standard procedure we can obtain the Lagrangian
$${\cal L} = - {1\over\sqrt{1-E^2}}\sqrt{{1\over1-E^2} -\cos^2\theta~\dot\phi^2}~\sin^3\theta + \sin^4\theta ~\dot\phi.\eqno{(3.6)}$$ 
After the calculations the momentum conjugates to $\phi$ becomes
$$P={\cos^2\theta \sin^3\theta~\dot\phi\over \sqrt{1 -\left(1-E^2\right)\cos^2\theta~\dot\phi^2}} + \sin^4\theta  ,\eqno{(3.7)}$$
and associated energy of the rotating deformed giant graviton is 
$$H = {P^2-2P\sin^4\theta + {\cos\theta \sin^6\theta\over\sqrt{1-E^2}}\sqrt{{1\over 1-E^2}-\sin^2\theta}~ +\sin^8\theta\over \sqrt{1-E^2}~\cos\theta~\sqrt {P^2-2P\sin^4\theta + {1\over 1-E^2}\sin^6}},\eqno{(3.8)}$$
which becomes that in the undeformed system in the limit of $E =0$.

    We have used above relation to perform many numerical evaluations and see that  the deformed giant graviton in the electric Melvin background, if it exist, will always have larger energy than the point-like graviton.   Thus the electric Melvin field will render the giant graviton unstable.

   It is interesting to see that  this property could also be found in analyzing the case of small electric, $E^2 \ll 1$.   In this limit the energy of the deformed giant graviton (3.8) has a simple form
$$H \approx {\sqrt{P^2-2P\sin^4\theta +\sin^6\theta}\over \cos\theta}+ {E^2\over2\cos\theta}{P^2-2P\sin^4\theta +2\sin^6\theta - \sin^8\theta\over \sqrt{P^2-2P\sin^4\theta +\sin^6\theta}}.\eqno{(3.9)}$$
The radius calculated from the above equation is
$$R_{gg}^2 \equiv \sin^2\theta \approx P + {3E^2\over 2}\left(P^2 - P\right).\eqno{(3.10)}$$
The energy of the above giant graviton $H(R_{gg})$ and that of the point giant graviton $H(0)$ calculated from (3.9) have the following relation
$$H(0) \approx  P +{E^2\over2}P~,~~~H(R_{gg})\approx P +{E^2\over2}\left(P+P^2 \right),~~\Rightarrow~~
H(R_{gg})-H(0) \approx {E^2\over 2} P^2.\eqno{(3.11)}$$
Thus the electric-field deformed configuration has larger energy than the trivial graviton and there does not exist stable giant graviton under the electric Melvin flux.

 Note that we have only considered the giant graviton on the deformed $S^5$ spacetimes in this paper.  The coordinates of the deformed $AdS_5$ spacetime in (2.1) or (3.1) are mathematically complex and it is difficult to study the ``dual'' giant graviton [3] therein.   The vibration modes of above deformed giant gravitons, which can improve our understanding of the holographic dual [5], are also difficult to find due to the mathematical complex.   The problems remain to be investigated.

\section{Conclusion}
It is well-known that a collection of p-branes in a background field strength can undergo an expansion into a higher-dimensional p+2 spherical brane. This is the so-called dielectric effect [25].  The new branes are stable due to a dynamical equilibrium between the tension of the spherical branes that makes them contract and the momentum in an external field that makes them expand.  The giant graviton, being a BPS object, is also formed from the dielectric effect [1,3].

  The previous investigations of the giant gravitons in the background B-field [11] or in Frolov's three parameter generalization of the Lunin-Maldacena background [13,14] had shown that the deformations have inclinations to make the giant graviton unstable.  In previous paper [15] we constructed the electric/magnetic Melvin spacetimes and studied the classical rotating string in these geometries.  Results therein showed that the external electric/magnetic field have inclinations to make these classical rotating string unstable.  In this paper, we study the giant graviton on the electric/magnetic field deformed Melvin geometries and see that, while the magnetic Melvin field has inclination to stabilize the giant graviton, the external electric Melvin filed will always render the giant graviton unstable.  We also see that while increasing the angular momentum of giant graviton the radius of the deformed giant graviton is initially an increasing function, the property as that in the undeformed giant graviton.   However, after it reach its maximum value the radius of the giant graviton becomes a decreasing function of the angular momentum.  During these regions the giant graviton is still a stable configuration, contrast to that in the undeformed theory.   Finally, beyond a critical value of angular momentum $P_c$ the configuration has higher energy than that of the point-like graviton and giant graviton  eventually becomes unstable.   We also find that the critical value of the angular momentum is an increasing function of the magnetic field flux .

  Finally, as giant gravitons can also take place in backgrounds different from AdS $\times$ S [9,10], it is interesting to see whether the properties found in this paper would also be shown in other electric/magnetic field deformed geometries.   The problem remains to be studied.
~

~

~

{\bf  \Large REFERENCES}
{\small
\begin{enumerate}
\item  J. McGreevy, L. Susskind and N. Toumbas, ``Invasion of Giant Gravitons from Anti de Sitter Space'', JHEP  0006 (2000) 008 [hep-th/0003075].
\item  J. Maldacena and A. Strominger, ``AdS3 Black Holes and a Stringy Exclusion Principle'', JHEP 9812 (1998) 005 [hep-th/9804085].
\item  M. T. Grisaru, R. C. Myers and O. Tafjord, ``SUSY and Goliath'', JHEP  0008 (2000) 040 [hep-th/0008015]; A. Hashimoto, S. Hirano and N. Itzakhi, ``Large Branes in AdS and their Field Theory Dual," 0008 (2000) 051 (2000)  [hep-th/0008016].
\item S. Das, A. Jevicki and S. Mathur, ``Giant Gravitons, BPS Bounds and 
Noncommutativity," Phys. Rev. D63 (2001) 044001 [hep-th/0008088].
\item S. Das, A. Jevicki and S. Mathur, `` Vibration Modes of Giant Gravitons," Phys. Rev. D63 (2001) 024013 [hep-th/0009019]; P. Ouyang``Semiclassical Quantization of Giant Gravitons ,'' [hep-th/0212228].
\item  J. Lee, ``Tunneling between the giant gravitons in AdS5 x S5'', Phys.Rev. D64 (2001) 046012 [hep-th/0010191].
\item B. Janssen and Y. Lozano,``A Microscopical Description of Giant Gravitons,'' Nucl.Phys. B658 (2003) 281 [hep-th/0207199].
\item B. Janssen and Y. Lozano, ``Non-Abelian Giant Gravitons,'' Class. Quant. Grav. 20 (2003) S517 [hep-th/0212257]. 
\item  S. R. Das, S. P.. Trivedi, S. Vaidya,``Magnetic Moments of Branes and Giant Gravitons,''  JHEP 0010 (2000) 037 [hep-th/0008203].
\item  J. M. Camino,``Worldvolume Dynamics of Branes,'' [hep-th/0210249]. 
\item  J. M. Camino and A.V. Ramallo, ``Giant gravitons with NSNS B field," JHEP 0109 (2001) 012 [hep-th/0107142];  ``M-Theory Giant Gravitons with C field," Phys.Lett. B525 (2002) 337 [hep-th/0110096 ]; S. Prokushkin and M. M.  Sheikh-Jabbari, ``Squashed giants: Bound states of giant gravitons,'' JHEP 0407 (2004) 077 (2004) [hep-th/0406053]. 
\item A. Mikhailov, ``Giant gravitons from holomorphic," JHEP 0011(2000) 027 [hep-th/0010206]; ``Nonspherical giant gravitons and matrix theory," [hep-0208077].
\item O. Lunin and J.M. Maldacena, ``Deforming Field Theories with $U(1)\times U(1)$ global symmetry and their gravity duals," JHEP 0505 (2005) 033 [hep-th/0502086]; S.A. Frolov, ``Lax Pair for Strings in Lunin-Maldacena Background," JHEP 0505 (2005) 069 [hep-th/0503201]; L. F. Alday, G. Arutyunov, S. Frolov, ``Green-Schwarz Strings in TsT-transformed backgrounds'' [hep-th/0512253]
\item R. de M. Koch, N. Ives, J. Smolic, M. Smolic, ``Unstable Giants," [hep-hep-th/0509007].
\item Wung-Hong Huang, ``Semiclassical Rotating Strings in  Electric and Magnetic Fields Deformed $AdS_5 \times S^5$ Spacetime'', Phys. Rev. D73 (2006) 026007 [hep-th/0512117].  
\item  Wung-Hong Huang, ``Thermal Instability of Giant Graviton in Matrix Model on PP-wave Background," Phys.Rev. D69 (2004) 067701 [hep-th/0310212]; H. Shin and, K. Yoshida, ``Thermodynamics of Fuzzy Spheres in PP-wave Matrix Model," Nucl.Phys. B701 (2004) 380 [hep-th/0401014 ];
N. Kawahara, J. Nishimura and K. Yoshida,``Dynamical aspects of the plane-wave matrix model at finite temperature,'' [hep-th/0601170 ].
\item J.~M.~Maldacena, ``The large N limit of superconformal field theories and supergravity,'' Adv. Theor. Math.  Phys. 2 (1998) 231 [hep-th/9711200];
 S. S. Gubser, I. R. Klebanov and A.~M.~Polyakov, ``Gauge theory correlators from non-critical string theory,'' Phys. Lett. B428 (1998) 105 [hep-th/9802109]; E.~Witten, ``Anti-de Sitter space and holography,'' Adv. Theor. Math. Phys. 2 (1998) 253 [hep-th/9802150].
\item  V. Balasubramanian, M. Berkooz, A. Naqvi and M. Strassler", Giant Gravitons in Conformal Field Theory", JHEP 0204 (2002) 034 [hep-th/0107119];
S. Corley, A. Jevicki and S. Ramgoolam, ``Exact Correlators of Giant Gravitons from Dual N = 4 SYM Theory,'' Adv. Theor. Math. Phys. 5 (2002) 809 
2002  [hep-th/0111222].
\item  S. Corley and S. Ramgoolam, ``Finite Factorization equations and sum rules for BPS Correlators in N=4 SYM Theory,'', Nucl. Phys. B641 (2002) 131  [hep-th/0205221];  D. Berenstein, ``Shape and Holography: Studies of dual operators to giant gravitons," Nucl. Phys. B675 (2003) 179 [hep-th/0306090];  R. de M. Koch and R. Gwyn, ``Giant Graviton Correlators from Dual $SU(N)$ super Yang-Mills Theory," JHEP 0411 (2004) 081 [hep-th/0410236].
\item R. Roiban, ``On spin chains and field theories,'' JHEP 0409 (2004) 023 [hep-th/0312218]; D. Berenstein and  S. A. Cherkis,``Deformations of N=4 SYM and integrable spin chain models,'' Nucl.Phys. B702 (2004) 49 [hep-th/0405215]
; S.A. Frolov, R. Roiban, A.A. Tseytlin, ``Gauge-string duality for superconformal deformations of N=4 Super Yang-Mills theory,'' JHEP 0507 (2005) 045 [hep-th/0503192].
\item  C.~G.~Callan, J.~A.~Harvey and A.~Strominger, ``Supersymmetric string solitons,'' [hep-th/9112030]; A.~Dabholkar, G.~W.~Gibbons, J.~A.~Harvey and F.~Ruiz Ruiz, ``Superstrings and Solitons,'' Nucl.\ Phys.\ B  340 (1990) 33; G.T. Horowitz and A.~Strominger, ``Black strings and P-branes,'' Nucl.\ Phys.\ B  360 (1991) 197.
\item M.A. Melvin, ``Pure magnetic and electric geons,'' Phys. Lett. 8 (1964) 65; F. Dowker, J. P. Gauntlett, D. A. Kastor and Jennie Traschen, ``Pair Creation of Dilaton Black Holes,'' Phys.Rev. D49 (1994) 2909-2917  [hep-th/9309075]; F.~Dowker, J.~P.~Gauntlett, D.~A.~Kastor and J.~Traschen, ``The decay of magnetic fields in Kaluza-Klein theory,'' Phys.\ Rev.\ D52 (1995) 6929 [hep-th/9507143]; M.~S.~Costa and M.~Gutperle, ``The Kaluza-Klein Melvin solution in M-theory,'' JHEP 0103 (2001) 027 [hep-th/0012072].
\item G.~W.~Gibbons and D.~L.~Wiltshire, ``Space-time as a membrane in higher dimensions,'' Nucl.\ Phys.\ B287 (1987) 717 [hep-th/0109093]; G.~W.~Gibbons and K.~Maeda, ``Black holes and membranes in higher dimensional theories with dilaton fields,'' Nucl.\ Phys.\ B298 (1988) 741; L. Cornalba and M.S. Costa, ``A New Cosmological Scenario in String Theory,'' Phys.Rev. D66 (2002) 066001 [hep-th/0203031]; T. Friedmann and H. Verlinde, ``Schwinger pair creation of Kaluza-Klein particles: Pair creation without tunneling,'' Phys.Rev. D71 (2005) 064018 [hep-th/0212163].
\item P. Ginsparg and C. Vafa, Nucl. Phys. B289 (1987) 414; T. Buscher, Phys. Lett. B159 (1985) 127; B194 (1987) 59; B201 (1988) 466; S. F. Hassan, ``T-Duality, Space-time Spinors and R-R Fields in Curved Backgrounds,'' Nucl.Phys. B568 (2000) 145 [hep-th/9907152].
\item R. Emparan, ``Born-Infeld Strings Tunneling to D-branes,'' Phys.Lett. B423 (1998) 71 [hep-th/9711106]; R. C. Myers, ``Dielectric-Branes,'' JHEP 9912 (1999) 022 [hep-th/9910053].
\end{enumerate}}
\end{document}